\documentclass[11pt]{article}
\usepackage{a4p}
\usepackage{amssymb}
\usepackage{epsfig}
\usepackage{cite}
\usepackage{pennames}
\def\ii{\'{\i}}
\def\nostrocostruttino#1\over#2{\mathrel{\mathop{\kern 0pt \rlap
{\hbox{$#1$}}} \hbox{\kern-.135em $#2$}}}

\newcommand{\ZP}[1]{{\it Z.\ Phys.}\ {\bf #1}}
\newcommand{\PL}[1]{{\it Phys.\ Lett.}\ {\bf #1}}
\newcommand{\PR}[1]{{\it Phys.\ Rev.}\ {\bf #1}}

\newcommand{\beq}{\begin{equation}}
\newcommand{\eeq}{\end{equation}}
\newcommand{\barr}{\begin{eqnarray}}
\newcommand{\earr}{\end{eqnarray}}
\newcommand{\ba}{\begin{array}}
\newcommand{\ea}{\end{array}}
 \def\Pb{\relax\ifmmode{\rm b}%
         \else${\rm b}$\fi}%
 \def\Ph{\relax\ifmmode{\rm h}%
         \else${\rm h}$\fi}%
 \def\PX{\relax\ifmmode{\rm X}%
         \else${\rm X}$\fi}%

\def\lsim{\mathrel{\rlap{\lower4pt\hbox{\hskip1pt$\sim$}}\raise1pt\hbox{$<$}}} 
\def\gsim{\mathrel{\rlap{\lower4pt\hbox{\hskip1pt$\sim$}}\raise1pt\hbox{$>$}}}

\newcommand{\la}{\lambda}
\begin{document}
\begin{titlepage}
 \begin{center}{\large   EUROPEAN ORGANISATION FOR NUCLEAR RESEARCH
 }\end{center}\bigskip
\begin{flushright}
  CERN-PPE/97-136 \\ 1$^{st}$ October 1997
\end{flushright}
\bigskip\bigskip\bigskip\bigskip\bigskip
\begin{center}
 {\huge\bf \boldmath
Quark fragmentation into vector and pseudoscalar mesons at LEP
 }
\end{center}
%\begin{flushright}
%{\bf OPAL IP063}\\
%DFTT 60/97 \\
%%INFNCA-TH9707 \\
%%hep-ph/9710505 \\
%\end{flushright}
%\vskip 1.5cm
%\begin{center}
%{\bf 
%Quark fragmentation into vector and pseudoscalar mesons at LEP
%}\\
\vskip 1.5cm
\begin{center}
{\sf M. Anselmino$^1$, M. Bertini$^{1,2}$, C. Burgard$^3$, 
F. Caruso$^4$ and P. Quintairos$^{1,4}$}
\vskip 0.8cm
{$^1$Dipartimento di Fisica Teorica, Universit\`a di Torino and \\
      INFN, Sezione di Torino, Via P. Giuria 1, 10125 Torino, Italy\\
\vskip 0.5cm
$^2$Institut de Physique Nucl\'eaire de Lyon \\
43 Bvd. du 11 Novembre 1918, F-69622 Villeurbanne Cedex \\
\vskip 0.5cm
$^3$CERN, PPE Division, CH-1211 Geneva 23, Switzerland \\
\vskip 0.5cm
$^4$Centro Brasileiro de Pesquisas F\ii sicas \\
R. Dr. Xavier Sigaud 150, 22290-180, Rio de Janeiro, Brazil} \\

\end{center}
\vskip 1.5cm
\noindent
{\bf Abstract:}

\vspace{6pt}

\noindent
Some data on the ratio of vector to vector + pseudoscalar mesons, 
$V/(V+P)$, 
and the probability of helicity zero vector states, $\rho_{00}(V)$, are now 
available from LEP. A possible relation between these two quantities and 
their interpretation in terms of polarized fragmentation functions are 
discussed; numerical estimates are given for the relative occupancies of 
$\PK$ and $\PK^{*}$, $\PD$ and $\PD^*$, $\PB$ and $\PB^*$ states. 

\vskip 2.0cm
\begin{center}
{\it Submitted to Phys. Lett. B}
\end{center}
\end{titlepage}
%%%%%%%%%%%%%%%%%%%%%%%%%%%%%%%%%%%%%%%%%%%%%%%%%%%%%%%%%%%%%%%%%%%%%%%%%%%%%%%
% \newpage
%%%%%%%%%%%%%%%%%%%%%%%%%%%%%%%%%%%%%%%%%%%%%%%%%%%%%%%%%%%%%%%%%%%%%%%%%%%%%%%
\pagestyle{plain}
\setcounter{page}{1}
\noindent
{\bf 1 - Introduction}
\vskip 12pt
The inclusive production of hadrons in 
$\Pe^+\Pe^-$ annihilations at LEP, 
$\Pe^+\Pe^- \to \Ph + \PX$, 
allows the fragmentation 
properties of quarks to be studied; 
this fragmentation process has recently received much attention both 
theoretically \cite{noi}--\cite{binn} and experimentally \cite{burg}, 
\cite{data}. It yields information on basic non-perturbative aspects of 
strong interactions or, in the case of heavy quarks, may test existing 
perturbative computations of fragmentation functions. 

We consider here some spin dependence of particular fragmentation processes,
namely the production of vector and pseudoscalar mesons,
their relative abundance and the probability for the vector meson
to be in a zero helicity state. These quantities can be related to polarized
and unpolarized quark fragmentation functions and their measurement
supplies basic information on the hadronization process of a quark; 
simple statistical models for such a process are discussed.

The quarks produced in $\Pe^+\Pe^-$ 
annihilation at LEP are longitudinally 
polarized with a probability $\rho_{\pm \pm}(\Pq)$ of being in 
a $\pm$ helicity state; the off-diagonal elements of the quark helicity 
density matrix $\rho_{\pm \mp}(\Pq)$ can be neglected. The 
probability for a hadron \Ph\ produced in the fragmentation of such a quark 
\Pq\ to be in a helicity $\la^{\,}_\Ph$ state is then given by
\beq
\rho^{\,}_{\la^{\,}_\Ph \la^{\,}_\Ph}(\Ph)
= {1\over \sum_{\,\Pq} \, D_\Pq^\Ph} \> \sum_\Pq \left[
D^{\Ph, \la^{\,}_\Ph}_{\Pq,+} \> \rho_{++}(\Pq)
+ D^{\Ph, \la^{\,}_\Ph}_{\Pq,-} \> \rho_{--}(\Pq) \right] \,,
\label{rholh}
\eeq 
where $D^{\Ph, \la^{\,}_\Ph}_{\Pq, \la^{\,}_\Pq}$ is the polarized 
fragmentation function of a quark \Pq\ with helicity $\la^{\,}_\Pq$ into 
a hadron \Ph\ with helicity $\la^{\,}_\Ph$. The unpolarized fragmentation
function is 
\beq
D^\Ph_\Pq = {1 \over 2} \sum_{\la^{\,}_\Ph, \la^{\,}_\Pq}
D^{\Ph, \la^{\,}_\Ph}_{\Pq, \la^{\,}_\Pq} = 
\sum_{\la^{\,}_\Ph} D^{\Ph, \la^{\,}_\Ph}_\Pq \,.
\label{dunp}
\eeq 

Equation (\ref{rholh}) holds in a probabilistic scheme, with a single-quark 
independent fragmentation. In Refs. \cite{akp} and \cite{aamr} it has been 
shown that for diagonal matrix elements this is a good approximation and 
that the coherent fragmentation of the quark, which takes into account its 
interaction with the $\bar \Pq$, only induces corrections of 
\cal{O}$(p_{\mathrm t}/(x\sqrt s))^2$, where $p_{\mathrm t}$ 
is the transverse momentum of 
the hadron relative to the jet axis and $x$ is the fraction of the quark 
energy carried by the hadron.

Let us first consider final hadrons which predominantly originate from
the fragmentation of only one type of quark, like mesons containing one
heavy flavour, $\PD,\PD^*$, $\PB,\PB^*$ and possibly $\PK,\PK^*$ at large $x$. 
In such a case one obtains
\barr
\rho_{11} &=& {1 \over D^V_\Pq} \left[ D^{V,1}_{\Pq,+} \> \rho_{++}(\Pq)
+ D^{V,1}_{\Pq,-} \> \rho_{--}(\Pq) \right] \,, \nonumber \\
\rho_{00} &=& {1 \over D^V_\Pq} \> D^{V,0}_\Pq \,, \label{rho} \\
\rho_{-1-1} &=& {1 \over D^V_\Pq} \left[ D^{V,-1}_{\Pq,+} \> \rho_{++}(\Pq)
+ D^{V,-1}_{\Pq,-} \> \rho_{--}(\Pq) \right] \nonumber 
\earr
and
\beq
P_V \equiv {V \over V+P} = {D^V_\Pq \over D^V_\Pq + D^P_\Pq} \,,
\label{vvp}
\eeq
where $V$ stands for vector and $P$ for pseudoscalar mesons. Both 
above and in the following we exploit the parity invariance of the 
fragmentation process, i.e.
\barr
D^{V,0}_{\Pq,+} &=& D^{V,0}_{\Pq,-} \>\> = \>\> D^{V,0}_\Pq , \nonumber \\
D^{V,1}_{\Pq,+} + D^{V,-1}_{\Pq,+} &=& D^{V,1}_{\Pq,-} + 
D^{V,-1}_{\Pq,-} \>\> = 
\>\> D^{V,1}_\Pq + D^{V,-1}_\Pq \,.
\label{par}
\earr
Note that for valence quarks and meson wave functions with zero orbital 
angular momentum one expects
\beq
D^{V,1}_{\Pq,-} = D^{V,-1}_{\Pq,+} = 0 \,.
\label{ass}
\eeq

Experimentally, only $\rho_{00}(V)$ --- 
via observation of the angular distribution of 
the vector meson two-body decays --- and the ratio $V/(P+V)$ can be 
determined; 
let us discuss what can be learnt about the quark hadronization process 
from such measurements. These two observables do not depend on the quark 
polarization state, but only on the spin and helicity of the 
final meson.

We define
\beq
{D^{V,1}_\Pq + D^{V,-1}_\Pq \over D^{V,0}_\Pq} \equiv \gamma_\Pq^V \,;
\quad\quad\quad {D^P_\Pq \over D^{V,0}_\Pq} \equiv \beta_\Pq^V \,.
\label{def}
\eeq
Such quantities might depend on $x$. From equations 
(\ref{dunp})--(\ref{par}) we 
obtain
\beq
\rho_{00}(V) = {1 \over 1 + \gamma_\Pq^V}
\label{rho00}
\eeq
and 
\beq
P_V = {1 + \gamma_\Pq^V \over 1 + \gamma_\Pq^V + \beta_\Pq^V} \,,
\label{vvp2}
\eeq
which implies
\beq
P_V = {1 \over 1 + \beta_\Pq^V \> \rho_{00}} \, \cdot
\label{rel}
\eeq

Note that in pure statistical spin counting one finds 
\beq
\gamma_\Pq^V = 2  \,, \quad\quad\quad \beta_\Pq^V = 1 \,,
\label{stat}
\eeq
so that
\beq
\rho_{00} = {1 \over 3} \,, \quad\quad\quad P_V = {3 \over 4} \,,
\label{stat2}
\eeq
and the vector meson alignment
\beq
A = {1 \over 2} \> (3 \rho_{00} - 1)
\label{ali}
\eeq
is zero. 

Equations 
(\ref{rho00}) and (\ref{vvp2}) show the relation between the measurable
quantities $\rho_{00}(V)$ and $P_V$ and the parameters $\gamma_\Pq^V$
and $\beta_\Pq^V$, whose physical meaning is clear through the definitions
(\ref{def}). Equation (\ref{rel}) gives a relation, 
in terms of the parameter $\beta_\Pq^V$, between the probability for 
a quark to fragment into a vector meson and the probability for the vector 
meson to have zero helicity. Note
that an increase of $\rho_{00}$ implies a decrease of $P_V$ and vice versa
for a fixed value of $\beta_\Pq^V$. 

There are some models in the literature which predict a definite 
relation between $P_V$ and $\rho_{00}$. In Ref. \cite{fp}
one assumes $\beta = 1$ so that $P_V = 1/(1 + \rho_{00})$,
while in Ref. \cite{don} one assumes $\gamma = 1 + \beta$ and 
$P_V = 1/(2 - 2\rho_{00})$. 

In case of more than one flavour contributing to the final mesons, as 
for $\pi,\rho$, the expressions of $\rho_{00}$ and $P_V$ become
\beq
\rho_{00}(V) = {\sum_{\,\Pq} \, D^{V,0}_\Pq \over 
\sum_{\,\Pq} \, D^{V,0}_\Pq \> (1 + \gamma_\Pq^V)}
\label{rho00'}
\eeq
and 
\beq
P_V = {\sum_{\,\Pq} \, D^{V,0}_\Pq \> (1 + \gamma_\Pq^V) \over 
\sum_{\,\Pq} \, D^{V,0}_\Pq \> (1 + \gamma_\Pq^V + \beta_\Pq^V)} \,,
\label{vvp2'}
\eeq
while equation (\ref{rel}) changes to 
\beq
P_V = {\sum_{\,\Pq} \, D^{V,0}_\Pq \over 
\sum_{\,\Pq} \, D^{V,0}_\Pq \> (1 + \beta_\Pq^V \> \rho_{00})} \, \cdot
\label{rel'}
\eeq
This reduces to the simple results (\ref{rho00})--(\ref{rel}) not only in 
the case 
of a single flavour contributing, but also when 
$\gamma_\Pq^V$ and $\beta_\Pq^V$ do not depend on the flavour $\Pq$. 
This might be the case for valence quarks. 

\vskip 12pt
\noindent
{\bf 2 - Numerical values and their interpretation}
\vskip 12pt
To estimate the numerical values of the parameters $\gamma$ and $\beta$ we 
use data on $\PK,\PK^{*}$, $\PD,\PD^{*}$ and $\PB,\PB^*$ 
production obtained by the 
OPAL Collaboration \cite{data}. 
\barr 
\rho_{00}(\PK^*) &=& 0.550 \pm 0.050, \quad
P_V(\PK) = 0.750 \pm 0.102,
\quad {\mathrm for}~~x > 0.5
\label{data1} \\
\rho_{00}(\PD^*) &=& 0.400 \pm 0.020, \quad
P_V(\PD) = 0.570 \pm 0.060,
\quad {\mathrm for}~~x > 0.2
\label{data2} \\
\rho_{00}(\PB^*) &=& 0.360 \pm 0.092, \quad
P_V(\PB) = 0.760 \pm 0.090, 
\quad {\mathrm for}~~x > 0.33
\label{data3}
\earr
where $x$ is the energy fraction of the mesons. 
The contributions from weak decays were subtracted, 
where appropriate
\footnote{
We have interpolated the charged kaon rates to $x>0.5$ using a fit to 
the differential cross-section listed in Ref. \cite{data}. We subtracted the 
contributions from charm and bottom events to the production of strange 
mesons using the JETSET \cite{bib-jetset} model, assigning systematic 
errors to this subtraction as determined in Ref. \cite{afblight}.}.
This, via equations (\ref{rho00})--(\ref{rel}), yields
\barr
\gamma_s^{\PK^*} &=& 0.82 \pm 0.17 \,, \quad\quad\quad
\beta_s^{\PK^*} = 0.61 \pm 0.33 \,. \label{res1} \\
\gamma_c^{\PD^*} &=& 1.50 \pm 0.12 \,, \quad\quad\quad 
\beta_c^{\PD^*} = 1.89 \pm 0.47 \,. \label{res2} \\
\gamma_b^{\PB^*} &=& 1.78 \pm 0.71 \,, \quad\quad\quad
\beta_b^{\PB^*} = 0.88 \pm 0.49 \,. \label{res3}
\earr

The parameters $\gamma$ and $\beta$ have a natural definition in terms
of fragmentation functions, but the experimental results can also be 
interpreted in terms of the relative occupancies of all vector and 
pseudovector states \cite{burg}: let us denote by $P^{\la}_S$ the probability 
that the fragmenting quark produces a meson with spin $S$ and helicity 
$\la$ and let us consider only the production of vector and pseudovector 
mesons. Then we obtain
\beq
\rho_{00}(V) = {P_1^0 \over P_1^{\pm1} + P_1^0} 
\label{rho001} 
\eeq
and
\beq
P_V = P_1^{\pm1} + P_1^0
\label {vvp3} 
\eeq
with 
\beq
P_1^{\pm1} = P_1^1 + P_1^{-1} \,,  \quad\quad\quad 
P_1^{\pm1} + P_1^0 + P_0^0 = 1 \,.
\label{norm}
\eeq
In terms of previous parameters 
\beq
P_1^{\pm1} = {\gamma \over 1 + \gamma + \beta} \,, \quad\quad
P_1^0 = {1 \over 1 + \gamma + \beta} \,, \quad\quad
P_0^0 = {\beta \over 1 + \gamma + \beta} \,\cdot 
\label{new}
\eeq

The data (\ref{data1})--(\ref{data3}) give
\beq
P_1^{\pm1}(\PK^*) = 0.34 \pm 0.06 \quad\quad
P_1^0(\PK^*) = 0.41 \pm 0.07 \quad\quad
P_0^0(K) = 0.25 \pm 0.10 \label{res4} 
\eeq
\beq
P_1^{\pm1}(\PD^*) = 0.34 \pm 0.04 \quad\quad 
P_1^0(\PD^*) = 0.23 \pm 0.03 \quad\quad
P_0^0(\PD) = 0.43 \pm 0.06 \label{res5} 
\eeq
\beq
P_1^{\pm1}(\PB^*) = 0.49 \pm 0.09 \quad\quad
P_1^0(\PB^*) = 0.27 \pm 0.08 \quad\quad
P_0^0(\PB) = 0.24 \pm 0.09 \label{res6}
\eeq
in agreement with equations (\ref{res1})--(\ref{res3}) and (\ref{new}). Simple 
spin counting would give $P_1^{\pm1} = 0.5$ and $P_1^0 = P_0^0 = 0.25$.

\vskip 12pt
\noindent
{\bf 3 - Comments and conclusions} 
\vskip 12pt

We have shown how the simultaneous measurements of the ratio of vector 
to vector + pseudovector mesons and $\rho_{00}(V)$ supply
basic information on the fragmentation of quarks; such information
does not depend on the helicity of the quark, but on the spin and
helicity of the final meson. Some data are available for $\PK$, 
$\PD$ and $\PB$ mesons and in some cases show clear deviations from simple 
statistical spin counting and predictions of other models. Such
information could be of crucial importance for the correct formulation
of quark fragmentation Monte Carlo programs, which at the moment 
widely assume simple relative statistical probabilities. 

Let us consider our results, equations (\ref{res1})--(\ref{res3}) and 
(\ref{res4})--(\ref{res6}). 
For strange mesons the data agree with spin counting in the amount of
$\PK$ versus $\PK^*$. However, among vector mesons, helicity zero states seem
to be favoured; in absolute terms, these are the most abundantly produced,
$P_1^0(\PK^*) = 0.41$. For charmed mesons both values of $\gamma_c^{\PD^*}$
and $\beta_c^{\PD^*}$ differ from the spin counting values (\ref{stat}),
suggesting a prevalence of pseudoscalar states, $P_0^0(\PD) = 0.43$, and, 
among vector mesons, of helicity zero states. They also disagree with 
the models of Refs. \cite{fp} and \cite{don}, which, respectively,
predict $\beta = 1$ and $\gamma = 1 + \beta$.  
The heavy $\Pb$-mesons, instead, are produced in good agreement with 
statistical spin counting rules, as one expects.  

Data from LEP1 are still being analysed. For this kind of study it
is important that the production rates for the scalar and
vector mesons and the results on spin alignment 
are presented on the same footing, i.e. with the same cuts on the
scaled energies and a consistent subtraction of the contributions from 
weak decays. It will be interesting to determine the parameters 
$\gamma$ and $\beta$ and the relative spin state occupancies $P_S^\lambda$ 
for other mesons, like $\pi$ and $\rho$. There might be some common features 
to all light quark fragmentation processes, leading to the same vector and 
pseudovector state occupancy probabilities for $\pi,\rho$ and 
$\PK,\PK^*$. 

It would also be worthwhile to study the dependence of these probabilties 
on the scaled energy of the mesons.
Also, the degree of universality of quark fragmentation could be tested
by studying the same quantities discussed here in other 
processes, like meson production in lepton--hadron or photon--hadron 
interactions; or in $\gamma$--$\gamma$ and hadron--hadron collisions. 
It would also be interesting to compare data on the production
of spin 1/2 and spin 3/2 baryons. Some non perturbative aspects of 
strong interactions can only be tackled by gathering experimental 
information and looking for patterns and regularities which might allow 
the formulation of correct phenomenological models. 

%\newpage
\vskip 24pt
\noindent
{\bf Acknowledgements}
\vskip 6pt
We would like to thank George Lafferty for helpful discussions.
M. Bertini, F. Caruso and P. Quintairos would like to thank the Department 
of Theoretical Physics of the Universit\`a di Torino for its kind hospitality; 
F.Caruso is grateful to INFN and to CNPq of Brazil for financial support;
P.Quintairos is grateful to the CNPq.

\end{document}